\documentclass[fleqn,10pt]{wlscirep}
\usepackage[utf8]{inputenc}
\usepackage[T1]{fontenc}
\usepackage{multirow}
\usepackage{adjustbox}
\usepackage{makecell}
\title{Segmentation of glioblastomas in early post-operative multi-modal MRI with deep neural networks}

\author[1,2,+,*]{Ragnhild Holden Helland}
\author[3,4,+]{Alexandros Ferles}
\author[1]{André Pedersen}
\author[3,5]{Ivar Kommers}
\author[6]{Hilko Ardon}
\author[4,7]{Frederik Barkhof}
\author[8]{Lorenzo Bello}
\author[9]{Mitchel S. Berger}
\author[10]{Tora Dunås}
\author[8]{Marco Conti Nibali}
\author[11, 12]{Julia Furtner}
\author[9]{Shawn Hervey-Jumper}
\author[13]{Albert J. S. Idema}
\author[14]{Barbara Kiesel}
\author[15]{Rishi Nandoe Tewari}
\author[16]{Emmanuel Mandonnet}
\author[3,5]{Domenique M.J. Müller}
\author[17]{Pierre A. Robe}
\author[8]{Marco Rossi}
\author[18,19]{Lisa M. Sagberg}
\author[8]{Tommaso Sciortino}
\author[20]{Tom Aalders}
\author[21]{Michiel Wagemakers}
\author[14]{Georg Widhalm}
\author[22]{Marnix G. Witte}
\author[23]{Aeilko H. Zwinderman}
\author[18,24]{Paulina L. Majewska}
\author[10,25]{Asgeir S. Jakola}
\author[18,24]{Ole Solheim}
\author[3,5]{Philip C. De Witt Hamer}
\author[1,2]{Ingerid Reinertsen}
\author[3,5,+]{Roelant S. Eijgelaar}
\author[1,+]{David Bouget}

\affil[1]{Department of Health Research, SINTEF Digital, NO-7465 Trondheim, Norway}
\affil[2]{Department of Circulation and Medical Imaging, Norwegian University of Science and Technology, NO-7491 Trondheim, Norway}
\affil[3]{Cancer Center Amsterdam, Brain Tumor Center, Amsterdam University Medical Centers, 1081HV Amsterdam, The Netherlands}
\affil[4]{Department of Radiology and Nuclear Medicine, Amsterdam University Medical Centers, Vrije Universiteit, 1081 HV Amsterdam, The Netherlands}
\affil[5]{Department of Neurosurgery, Amsterdam University Medical Centers, Vrije Universiteit, 1081 HV Amsterdam, The Netherlands}
\affil[6]{Department of Neurosurgery, Twee Steden Hospital, 5042 AD Tilburg, The Netherlands}
\affil[7]{Institutes of Neurology and Healthcare Engineering, University College London, London WC1E 6BT, UK}
\affil[8]{Neurosurgical Oncology Unit, Department of Oncology and Hemato-oncology, Humanitas Research Hospital, Università Degli Studi di Milano,20122 Milano, Italy}
\affil[9]{Department of Neurological Surgery, University of California San Francisco, CA 94143, USA}
\affil[10]{Department of Clinical Neuroscience, Institute of Neuroscience and Physiology, Sahlgrenska Academy, University of Gothenburg, Gothenburg, Sweden}
\affil[11]{Department of Biomedical Imaging and image-guided Therapy, Medical University Vienna, 1090 Wien, Austria}
\affil[12]{Research Center for Medical Image Analysis and Artificial Intelligence (MIAAI), Faculty of Medicine and Dentistry, Danube Private University, 3500 Krems, Austria}
\affil[13]{Department of Neurosurgery, Northwest Clinics, 1815 JD Alkmaar, The Netherlands}
\affil[14]{Department of Neurosurgery, Medical University Vienna, 1090 Wien, Austria}
\affil[15]{Department of Neurosurgery, Haaglanden Medical Center, 2512 VA The Hague, The Netherlands}
\affil[16]{Hôpital Lariboisière, Department of Neurological Surgery, 75010 Paris, France}
\affil[17]{Department of Neurology and Neurosurgery, University Medical Center Utrecht, 3584 CX Utrecht, The Netherlands}
\affil[18]{Department of Neurosurgery, St. Olavs hospital, Trondheim University Hospital, NO-7030 Trondheim, Norway}
\affil[19]{Department of Public Health and Nursing, Norwegian University of Science and Technology, NO-7491 Trondheim, Norway}
\affil[20]{Department of Neurosurgery, Isala, 8025 AB Zwolle, The Netherlands}
\affil[21]{Department of Neurosurgery, University Medical Center Groningen, University of Groningen, 9713 GZ Groningen, The Netherlands}
\affil[22]{Department of Radiation Oncology, The Netherlands Cancer Institute, 1066 CX Amsterdam, The Netherlands}
\affil[23]{Department of Clinical Epidemiology and Biostatistics, Amsterdam University Medical Centers, University of Amsterdam, 1105 AZ Amsterdam, The Netherlands}
\affil[24]{Department of Neuromedicine and Movement Science, Norwegian University of Science and Technology, NO-7491 Trondheim, Norway}
\affil[25]{Department of Neurosurgery, Sahlgrenska University Hospital, Gothenburg, Sweden}
\affil[*]{ragnhild.h.helland@ntnu.no, Professor Brochs gate 2, 7030 Trondheim, Norway}

\affil[+]{these authors contributed equally to this work}

\keywords{Segmentation, Glioblastoma, Post-operative MRI, Deep Learning}

\begin{abstract}
Extent of resection after surgery is one of the main prognostic factors for patients diagnosed with glioblastoma. To achieve this, accurate segmentation and classification of residual tumor from post-operative MR images is essential. The current standard method for estimating it is subject to high inter- and intra-rater variability, and an automated method for segmentation of residual tumor in early post-operative MRI could lead to a more accurate estimation of extent of resection. 
In this study, two state-of-the-art neural network architectures for pre-operative segmentation were trained for the task. The models were extensively validated on a multicenter dataset with nearly 1000 patients, from 12 hospitals in Europe and the United States.
The best performance achieved was a 61\% Dice score, and the best classification performance was about 80\% balanced accuracy, with a demonstrated ability to generalize across hospitals. 
In addition, the segmentation performance of the best models was on par with human expert raters. The predicted segmentations can be used to accurately classify the patients into those with residual tumor, and those with gross total resection. 
\end{abstract}
\begin{document}

\flushbottom
\maketitle
%
%
\thispagestyle{empty}
\section{Introduction}
Glioblastoma, the most common malignant primary brain cancer, requires a multidisciplinary treatment approach comprising maximum safe surgical resection, followed by concurrent radiation and chemotherapy~\cite{Davis2016Glioblastoma:Treatment}. Even so, median survival in unselected patients is only 12 months~\cite{Skaga2021Real-worldApply}. Due to the high invasiveness, a complete resection of all tumor cells is not possible. Still, extensive surgical resections are associated with longer survival\cite{Coburger2017ImpactImaging.}, but as surgically induced neurological impairment is associated with shorter survival~\cite{aabedi2022association}, extent of resection (EOR) and surgical strategy, for example resection or biopsy only, needs to be weighed up against risks in individual patients.

The EOR is calculated as the ratio between the surgically-removed and pre-operative tumor volume , which relies on an accurate segmentation of the tumor tissue in both pre- and post-operative MR scans.
In recent years, a large body of work has focused exclusively on automated segmentation of pre-operative glioblastoma, yet the task of residual tumor segmentation from early post-operative MRI (EPMR) has gained less attention from the research community.
In current practice, the residual tumor size is estimated manually through eye-balling~\cite{Berntsen2020VolumetricReports}, or using crude measures such as the bi-dimensional product of the largest axial diameter of the contrast enhancing residual tumor, according to the Response Assessment in Neuro-Oncology (RANO) criteria~\cite{Wen2010UpdatedGroup}. Manual volume segmentations are more sensitive but expertise-dependent and time-consuming, with high inter- and intra-rater variability~\cite{Berntsen2020VolumetricReports,Visser2019Inter-raterMRI}. An automated method for post-operative tumor volume segmentation from EPMR would therefore be beneficial.

Glioblastoma segmentation from pre-operative MR scans has received a lot of attention in the literature in recent years. Many contributions were motivated by the MICCAI Brain Tumor Segmentation (BraTS) Challenge~\cite{Menze2015TheBRATS}. With the emergence of fully convolutional neural networks (CNNs)~\cite{Shelhamer2017FullySegmentation}, deep learning-based approaches have nearly completely replaced more conventional methods in medical image segmentation~\cite{wang2022medical}. Variants of the U-Net architecture~\cite{Ronneberger2015U-net:Segmentation} have facilitated the basis-architecture for the majority of auto-segmentation algorithms, including DeepMedic~\cite{Kamnitsas2017EfficientSegmentation}, Attention U-Net~\cite{Schlemper2019AttentionImages}, and the recently established nnU-Net~\cite{Isensee2021NnU-Net:Segmentation}, with state-of-the-art results in several medical image segmentation benchmarks. The winning submissions in the BraTS challenge in 2021 and 2022 were an extension of the nnU-Net architecture~\cite{Luu2021ExtendingSegmentation}, and an ensemble of three state-of-the art architectures for medical image segmentation, comprising nnU-Net, DeepSeg, and DeepSCAN~\cite{Zeineldin2022MultimodalSolution}, respectively.
In the absence of a publicly available dataset for residual tumor segmentation from EPMR, the literature on this problem is sparse when compared to the pre-operative segmentation task. Semi-automatic methods, combining of one or several voxel- or shape-based image segmentation algorithm, have been proposed from intensity thresholding (e.g., Otsu and relative entropy)~\cite{otsu1979threshold, Cordova2014QuantitativeTrials, Odland2015VolumetricGrowth}, fuzzy algorithms~\cite{Cordova2014QuantitativeTrials}, Gaussian mixture model~\cite{Zeng2016SegmentationInjuries}, morphological operations~\cite{Odland2015VolumetricGrowth}, region-based active contours~\cite{Chow2014SemiautomatedMultiforme, Krivoshapkin2019AutomatedGlioblastoma}, the level set approach~\cite{Zhu2012Semi-AutomaticEvaluation, Chow2014SemiautomatedMultiforme, Krivoshapkin2019AutomatedGlioblastoma}, and CNNs~\cite{Dhara2018InteractiveFollow-up}. Unfortunately, these methods relied on user inputs, either by manual initialisation, or by interactive refinement of the resulting segmentation. They are therefore challenging to use in clinical practice, and in large datasets. In addition, all validation studies were solely performed on single-center local datasets, consisting of 15 to 37 patients, making if difficult to demonstrate the generalizability of the proposed methods.

Regarding fully automated approaches, Meier et al.~\cite{Meier2017AutomaticGlioblastoma} presented an automated method based on decision forests for residual tumor segmentation using EPMR from 19 patients. A more recent work by Ghaffari et al.~\cite{Ghaffari2022AutomatedImages} proposed to fine-tune a 3D densely connected U-Net, pre-trained on the BraTS20 dataset, on a local dataset of 15 post-operative glioblastomas. However, the MR scans were all acquired for radiation therapy planning and not within the recommended time frame to acquire EPMR scans, within 72 hours after surgical resection~\cite{Wen2010UpdatedGroup}.
Deep learning approaches have recently shown to outperform more traditional algorithms on most image segmentation tasks, including segmentation of pre-operative glioblastomas~\cite{Luu2021ExtendingSegmentation,Zeineldin2022MultimodalSolution}. The  utmost requirement is the number of included patients and the quality of the MR images comprising a study dataset. Preferably, the data should originate from different locations, to evaluate the ability of the trained models to generalize across different hospitals, scanners, or clinical practice.

In this work, we determine the performance of two CNN architectures to segment residual enhancing glioblastoma on early post-operative scans. The selected architectures are the nnU-Net, state-of-the-art for pre-operative glioblastoma segmentation, and AGU-Net, an architecture developed for pre-operative segmentation of brain tumors. These architectures have both demonstrated excellent performance on pre-operative segmentation in previous studies on pre-operative brain tumor segmentation~\cite{Bouget2021GlioblastomaTask,Bouget2021MeningiomaMechanisms,Bouget2022PreoperativeReporting}, and they exhibit different strengths and weaknesses. The automatic results are compared with manual segmentations, using different combinations of MRI scans in a large dataset consisting of paired pre- and early post-operative MRI scans from 956 patients in 12 medical centers in Europe and the United States. Extensive validation studies are presented to identify the best architecture configuration, quantify the performances and ability to generalize, and highlight potential relevance for use in clinical practice. Finally, the best performing models are made publicly available and integrated into the open software Raidionics~\cite{Bouget2022PreoperativeReporting}.

\section{Materials \& Method}

\subsection{Data}
\begin{figure}[!ht]
\centering
\includegraphics[scale=0.8]{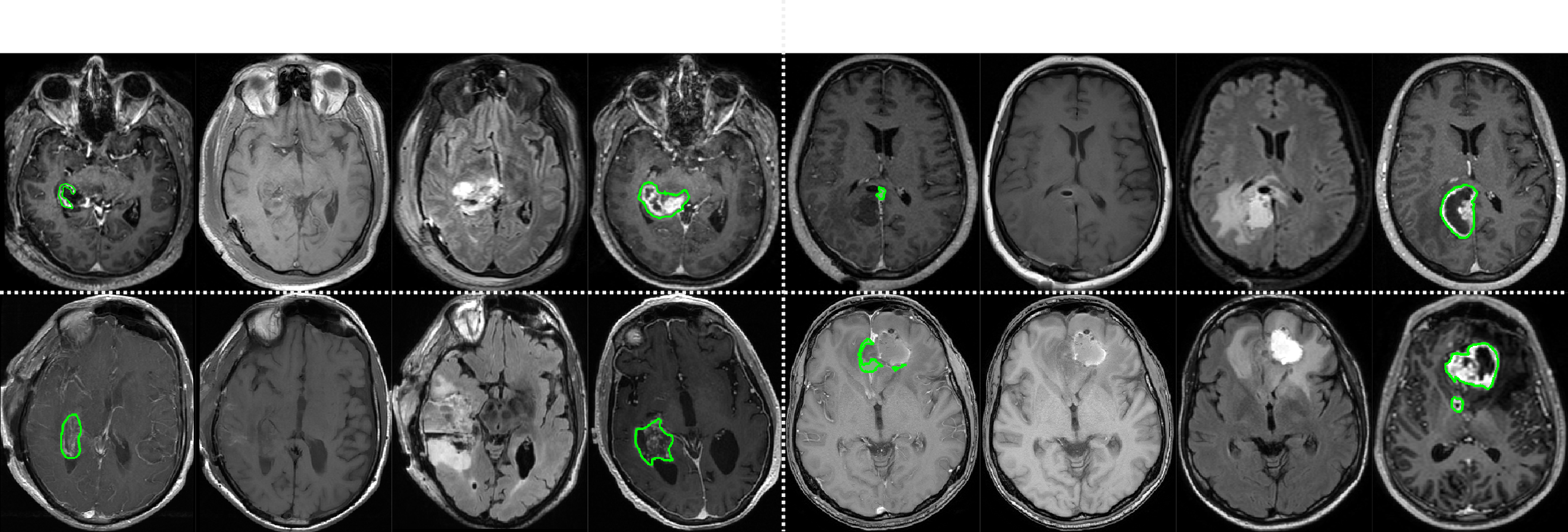}
\caption{Dataset examples for four patients, separated by white dash-lines. For each patient, an axial view from the EPMR T1w-CE, EPMR T1w, EPMR FLAIR, and pre-operative T1w-CE are displayed. Outlines of the manually annotated tumors are shown in green.}
\label{fig:dataset-illu}
\end{figure}

A dataset comprised of pre-operative and early post-operative MRI scans from 956 patients, who underwent surgical resection of glioblastoma, was assembled for this study. Twelve different hospitals across Europe and in the US contributed data, with the following patient distribution per center: 23 patients from the Northwest Clinics, Alkmaar, Netherlands (\textit{ALK}); 73 patients from the Amsterdam University Medical Centers, location VU medical center, Netherlands (\textit{AMS}); 43 patients from the University Medical Center Groningen, Netherlands (\textit{GRO}); 40 patients from the Medical Center Haaglanden, the Hague, Netherlands (\textit{HAG}); 55 patients from the Humanitas Research Hospital, Milano, Italy (\textit{MIL}); 41 patients from the H\^opital Lariboisi\`ere, Paris, France (\textit{PAR}); 108 patients from the University of California San Francisco Medical Center, U.S. (\textit{SFR}); 53 patients from the University Medical Center Utrecht, Netherlands (\textit{UTR}); 45 patients from the Medical University Vienna, Austria (\textit{VIE}); 51 patients from the Isala hospital, Zwolle, Netherlands (\textit{ZWO}); 237 patients from St. Olavs hospital, Trondheim University Hospital, Norway (\textit{STO}); and 187 patients from the Sahlgrenska University Hospital, Gothenburg, Sweden (\textit{GOT}).

The cohorts are subsets of a broader dataset, thoroughly described previously for their pre-operative content~\cite{Kommers2021GlioblastomaSegmentations}, for patients with available EPMR data.
All EPMR scans were acquired within 72 hours after surgery, with the exception of the UTR center where the limit used was up to one week post-surgery. The recommended time frame for acquiring the EPMR scans has been stated in the National Comprehensive Cancer Network (NCCN) recommendations\cite{Nabors2017CentralGuidelines}, in order to maximize differences between residual enhancing tumor and enhancement due to post-surgical changes in the tissue~\cite{Garcia-Ruiz2021PreciseGlioblastoma,Stupp2014High-gradeFollow-up}.
For each patient in the dataset, the following post-operative MRI sequences were acquired: T1-weighted (T1w), gadolinium-enhanced T1-weighted (T1w-CE), and T2-weighted fluid attenuated inversion recovery (FLAIR).

The residual tumor tissue was manually segmented in 3D in T1w-CE MR scans by trained annotators, supervised by expert neuroradiologists and neurosurgeons. The manual segmentation was performed using all available standard MR sequences, and residual tumor tissue was defined as enhancing tissue in the T1w-CE scan, but darker in the T1w scan. Hence, blood was distinguished from residual tumor by a hyperintense signal on T1w scans.
For each patient, a further post-operative distinction can be made between cases showcasing residual tumor (RT) in EPMR scans and cases presenting a gross total resection (GTR), defined as a residual tumor volume of less than 0.175\,ml.~\cite{Stummer2006Fluorescence-guidedTrial}. The cut-off was chosen to reduce risk of interpretation problems when distinguishing between tumour enhancement and that of non-specific enhancement, such as small vessels or enhancing pia mater. Under this paradigm, 352 patients (35\%) in our dataset had a GTR, whereas the remaining 604 patients had residual tumor. An overview of the data from the 12 hospitals is shown in Table~\ref{tab:data_distribution_hosp}, and some examples are illustrated in Figure~\ref{fig:dataset-illu}.

\begin{table}[ht]
 \caption{Dataset distributions and statistics across the twelve hospitals, represented by their acronyms. RT: residual tumor, GTR: gross total resection.}
  \centering
  \begin{tabular}{c|c|c|c|c|c|c|c|c|c|c|c|c}
    \textbf{Hospital} & \textbf{HAG} & \textbf{MIL} & \textbf{ZWO} & \textbf{VIE} & \textbf{ALK} & \textbf{PAR} & \textbf{SFR} & \textbf{GRO} & \textbf{UTR} & \textbf{AMS} & \textbf{STO} & \textbf{GOT}\tabularnewline
    \hline
    Patients & 40 & 55 & 51 & 45 & 23 & 41 & 108 & 43 & 53 & 73 & 237 & 187\tabularnewline
    \hline
    RT & 23 & 34 & 18 & 30 & 18 & 29 & 80 & 26 & 20 & 51 & 162 & 113\tabularnewline
    GTR & 17 & 21 & 33 & 15 & 5 & 12 & 28 & 17 & 33 & 22 & 75 & 74\tabularnewline
    \hline
    RT ratio (\%) & 57.5 & 61.8 & 35.3 & 66.7 & 78.3 & 70.7 & 74.1 & 60.5 & 37.7 & 69.9 & 68.4 & 60.4\tabularnewline
  \end{tabular}
  \label{tab:data_distribution_hosp}
\end{table}

In addition, 20 patients out of the 73 in the AMS cohort have been annotated eight times in total, by four novices raters and four experts raters. This cohort has been used in a previous study to evaluate the inter-rater variability of tumor annotation from annotators with different levels of experience~\cite{Visser2019Inter-raterMRI}, and will be referred to as the inter-rater variability dataset in the remainder of the document.

\subsection{Segmentation process}

\begin{figure}[!ht]
\centering
\includegraphics[scale=0.5]{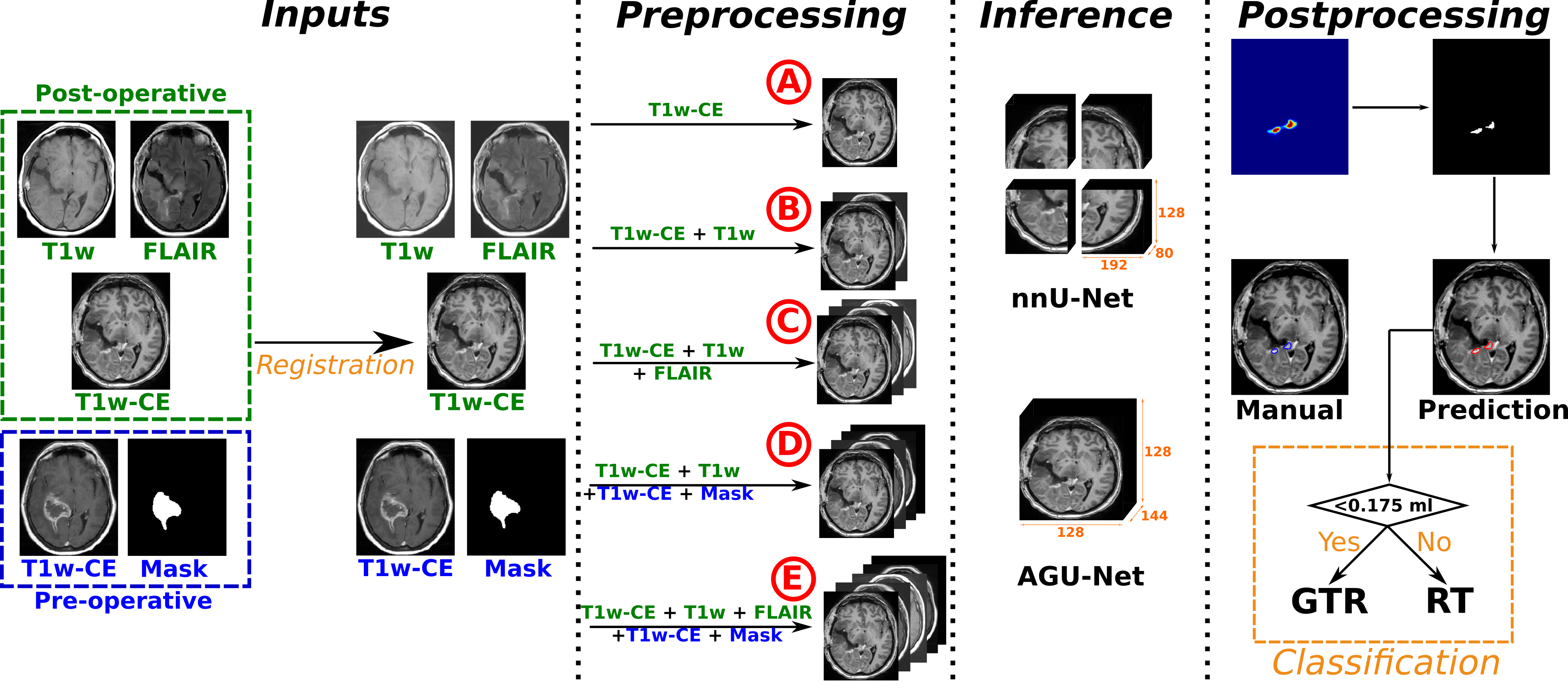}
\caption{Overall residual tumor segmentation pipeline from EPMR scans and classification between gross total resection or residual tumor. The registration is performed using the SyN approach from ANTs, multiple input configurations using different combinations of MR sequences were considered (noted from A to E), and two architectures were evaluated: nnU-Net and AGU-Net.}
\label{fig:pipeline-illu}
\end{figure}

Similar to our previous work on pre-operative glioblastoma segmentation~\cite{Bouget2021GlioblastomaTask}, the following two competitive CNN architectures were selected for the task of voxel-wise segmentation of residual tumor tissue: patch-wise nnU-Net~\cite{Isensee2021NnU-Net:Segmentation} and full-volume AGU-Net~\cite{Bouget2021MeningiomaMechanisms}.

Multiple MR sequences combinations can be considered as input for the CNN architectures. In an attempt to minimize essential input and following typical incremental assessment by neuroradiologists, these combinations of input sequences were considered for the automated segmentations of post-operative tumor: (\textbf{A}) the EPMR T1w-CE scan only, (\textbf{B}) the EPMR T1w-CE and EPMR T1w, to potentially distinguish between blood and residual tumor, (\textbf{C}) all standard EPMR sequences: T1w-CE, T1w, and FLAIR scans, (\textbf{D}) the EPMR T1w-CE and EPMR T1w, and the pre-operative T1w-CE MR scan and corresponding tumor segmentation mask, and (\textbf{E}) all standard EPMR sequences: T1w-CE, T1w, and FLAIR scans, and the pre-operative T1w-CE MR scan and corresponding tumor segmentation mask. An overview of the whole segmentation pipeline with the different input designs and subsequent steps is presented in the following sections, and illustrated in Fig.~\ref{fig:pipeline-illu}.

\subsubsection{Pre-processing} 
For proper anatomical consistency across the different inputs sequences, an initial image-to-image registration procedure was performed. The EPMR T1w-CE scan was elected as the reference space and all subsequent volumes were registered to it using the SyN diffeomorphic method~\cite{avants2008symmetric} from the Advanced Normalization Tools (ANTs) framework~\cite{avants2009advanced}. Skull-stripping was subsequently performed on all input MR scans, based on the brain mask from the EPMR T1w-CE scan. All brain masks were automatically generated using a pre-trained slab-wise AGU-Net model with input shape $256\times 192\times 32$\,voxels.
For the nnU-Net architecture, the pre-processing was automatically decided by the framework based on the dataset, and all inputs were resampled to $0.5\times 0.5\times 1.0\,\text{mm}^{3}$ spacing and zero-mean normalized. For the AGU-Net architecture, the full-resolution analysis required a lower resolution, and therefore all inputs were resampled to an isotropic $1.0\,\text{mm}^{3}$ spacing, resized to $128\times 128\times 144$\,voxels, and zero-mean normalized. 

\subsubsection{Training specifications for the nnU-Net architecture}

\textbf{Architecture design} From the nnU-Net framework analysis of the dataset, the 3D full-resolution U-Net with the following parameters was recommended, using $192\times 128\times 80$\,voxels as input patch size. The network used five levels, downsampling using strided convolution layers, and upsampling using transposed convolution layers. Kernel size of $1\times 3\times 3$\,voxels for the first level, $3\times 3\times 3$ for the remaining four levels, and filter sizes of $\{32,64,128,256,320\}$ were used for each level, respectively. The loss function was a combination of the Dice score and cross-entropy. A stride of one was used for the convolution layers.   

\textbf{Network training} All models were trained from scratch for 1000 epochs using a stochastic gradient descent with Nesterov momentum optimizer (momentum=0.99). One epoch was defined as 250 batch iterations with a batch size of two.
On-the-fly data augmentations were performed comprising rotation, scaling, additive Gaussian noise, Gaussian blur, brightness and contrast augmentation, and gamma augmentation.

\subsubsection{Training specifications for the AGU-Net architecture}
\label{subsec:agunet}
\textbf{Architecture design}
The AGU-Net, as described by Bouget et al.~\cite{Bouget2021MeningiomaMechanisms}, is a 3D U-Net architecture with an integrated attention-gated mechanism, with five block levels using filter sizes of \{16, 32, 128, 256, 256\}, respectively. The input size of the network was set to $128 \times 128 \times 144 \times S$, with $S$ being the number of sequences used as input. The architecture also uses multi-scale input and deep supervision. The class-averaged Dice loss, excluding the background, was used for training the different models.

\textbf{Network training}
All models were initialized using pre-trained weights from the best pre-operative glioblastoma segmentation model~\cite{Bouget2021GlioblastomaTask}, and only the input layer was adapted to account for the different input combinations considered.
The Adam optimizer was used with an initial learning rate of $1\times 10^{-3}$, and the training was stopped after 30 consecutive epochs without validation loss improvement. Gradient accumulation~\cite{pedersen2023gradacc} was performed to increase the batch size from 2 samples to 32, tackling graphics processing unit (GPU) memory limitations for large batch training. Data augmentation techniques were leveraged including horizontal and vertical flipping, random rotations in the range $[-20^{\circ}, 20^{\circ}]$, and a translation of up to 10\% of the axis dimension. Each augmentation was performed with a probability of 50\% for each training sample. 

\subsubsection{Post-processing and GTR classification}
During inference, residual tumor tissue was predicted by each trained model, resulting in a probability map of the same resolution as the EPMR T1w-CE scan.
A binary mask was then generated from the probability map, using the best threshold determined from the validation studies.
The binary mask was further refined by filtering out potential noise, inherent to the voxel-wise segmentation task, by applying a connected components analysis and removing any identified object smaller than 20\,voxels.
Finally, the refined binary mask was used to assess whether gross total resection has been achieved for the patient.  

\subsection{Validation studies}
\label{sec:validation_studies}
In this work, the trained models were assessed based on their ability to perform segmentation of the residual tumor and to classify patients into those with gross total resection and those with residual tumor.
For the segmentation task, only two classes are considered, whereby a positive voxel exhibits tumor tissue, whereas a negative voxel represents either background or normal tissue. For the classification task, a rest tumor volume threshold was selected to serve as cut-off value.

\subsubsection{Protocols}
The validation studies presented in this work were conducted following a five-fold cross-validation, summarized in Table~\ref{tab:data_distribution}.
First, all patients from 11 out of the 12 the hospitals in our dataset, excluding the AMS cohort, were split into five hospital-stratified folds, with an approximately balanced number of patients in each fold. The remaining 73 patients from the AMS hospital were kept as an hold-out test set. For each iteration of the cross-validation, four folds were used for training, the remaining fifth fold was used for validation, and the hold-out set was used for test.

This approach presents similar benefits to the leave-one-hospital-out strategy used in previous work~\cite{Bouget2021GlioblastomaTask}, with the advantage of a reduced training time. Finally, predictions over the test set were generated by ensembling over the predictions obtained by each of the five trained models. An average pooling voting scheme was applied to each of the model predictions, to produce a single softmax prediction.

\begin{table}[ht]
\caption{Distribution of hospitals and patient samples featured in the 5-fold validation sets and hold-out test set.}
  \centering
  \begin{tabular}{c|c|c|c|c|c|c}
    & \multicolumn{5}{c|}{\textbf{Hospital-wise cross-validation set}} & \multirow{2}{*}{\textbf{Hold-out set}}\tabularnewline
    \textbf{Fold} & \textbf{0} & \textbf{1} & \textbf{2} & \textbf{3} & \textbf{4} & \tabularnewline
    \hline
    Hospitals validation & STO & \makecell{GRO, MIL\\ UTR} & SFR, VIE & \makecell{PAR, ZWO\\ALK, HAG} & GOT & AMS\tabularnewline
    \hline
    Patients train & 646 & 732 & 730 & 728 & 696 & --- \tabularnewline
    Patients validation & 237 & 151 & 153 & 155 & 187 &73\tabularnewline
  \end{tabular}
  \label{tab:data_distribution}
\end{table}

\subsubsection{Metrics}
To evaluate the models' voxel-wise performance on the task of residual tumor segmentation, Dice scores were computed between the ground truth annotation and the post-processed binary prediction mask. The Dice scores are reported for the subgroup of patients with residual tumor tissue according to the ground truth annotation, labelled as the 'positive' (P) group. The Dice scores for the subgroup of patients with residual tumor according to the ground truth annotation and the network predictions, labelled as the 'true positive' (TP) group, are also reported. Pooled estimates, when computed from each fold's results, are reported for each measurement as mean and standard deviation (indicated by $\pm$) in the tables.

For the patient-wise classification task of distinguishing patients with gross total resection and patients with residual tumor, a standard sensitivity and specificity approach was conducted represented by the balanced accuracy score (noted bAcc). A residual tumor volume below the clinical volume threshold was thus counted as a negative (i.e., GTR) and as positive otherwise (i.e., RT). Following this consideration, a patient was considered a true positive (TP) if both the ground truth annotation residual tumor volume and detected residual tumor volume were $\geq 0.175$\,ml, for any given Dice score (i.e., $\geq 0.01$). Conversely, if both volumes were $<0.175$\,ml, the patient was labelled as a true negative (TN). Patients where the ground truth volume was above the threshold volume and the prediction was below were marked as false negatives (FN), and false positive (FP) vice versa.

In the case of inter-rater variability, the Jaccard score, closely related to the Dice score by \(J = \frac{D}{2-D}\), was used to compare the models' performance. The Jaccard was chosen for easy comparison with a previously published work on the same dataset~\cite{Visser2019Inter-raterMRI}.

\subsubsection{Experiments}
The following three experiments were conducted in this study:

(i) Residual tumor segmentation performance study: using the 5-fold cross-validation protocol and segmentation metrics, both nnU-Net and AGU-Net architectures' segmentation performances were compared for the five combinations of input sequences.

(ii) Gross total resection classification performance study: using the 5-fold cross-validation protocol, classification metrics, and best input combination identified in the first experiment, both architectures were compared in terms of ability to classify between gross total resection and residual tumor patients.

(iii) Inter-rater variability study: the best model from each architecture was benchmarked in terms of segmentation performance against the performance of novice and expert annotators, using the inter-rater variability dataset. For each patient, a consensus agreement annotation has been created using a majority voting approach. Using all eight annotations from both experts and novices, a voxel was defined as belonging to a tumor if annotated by more than half of the annotators. The models' binary predictions and the eight inter-rater annotations were then compared against the ground truth annotations (as used in the hold-out test set) and the consensus annotations.

\section{Results}
The studies were performed using multiple machines with the two following specifications: (i) Intel Core Processor (Broadwell, no TSX, IBRS) central processing unit (CPU) with 16 cores, 64GB of RAM, Tesla V100S (32GB) dedicated GPU, and a regular hard-drive and (ii) a GPU server with a total of 256 CPU cores, 2TB of RAM, and six NVIDIA A100-SXM4 (80GB) cards.
The AGU-Net architecture was implemented in Python 3.6 with the TensorFlow v1.13.1 library~\cite{abadi2015tensorflow}. For the nnU-Net architecture, Python 3.8, PyTorch v1.13.1~\cite{paszke2019pytorch}, and the nnU-Net framework v1.7.0~\cite{Isensee2021NnU-Net:Segmentation} were used.

\subsection{Residual tumor segmentation performance study}

\begin{table}[!ht]
 \caption{Segmentation performances for patients with residual tumor, for both architectures, all input configurations, and over the validation and test sets.}
  \centering
  {\adjustbox{max width=\textwidth}{
  \begin{tabular}{lll|cc|ccc}
    \multirow{2}{*}{\bfseries Input} & \multirow{2}{*}{\bfseries Prot.} & \multirow{2}{*}{\bfseries Arch.} & 
  \multicolumn{2}{c|}{\bfseries Voxel-wise} & \multicolumn{3}{c}{\bfseries Patient-wise} \\
       & & & \bfseries DSC-P & \bfseries DSC-TP & \bfseries Recall & \bfseries Precision & \bfseries F1 \tabularnewline
    \hline 
\multirow{4}{*}{A} & \multirow{2}{*}{Val} & nnU-Net & 46.94$\pm$24.03 & 49.51$\pm$21.70 & 99.81$\pm$0.35 & 62.96$\pm$6.60 & 77.03$\pm$4.98\tabularnewline 
 &  & AGU-Net & 37.72$\pm$29.54 & 51.05$\pm$22.28 & 79.70$\pm$6.69 & 80.75$\pm$5.79 & 79.83$\pm$3.01\tabularnewline 
 & \multirow{2}{*}{Test} & nnU-Net & 52.38$\pm$21.14 & 53.43$\pm$19.77 & 100.00 & 70.83 & 82.93\tabularnewline 
 &  & AGU-Net & 38.06$\pm$27.45 & 46.21$\pm$22.80 & 84.31 & 84.31 & 84.31\tabularnewline 
\hline 
\multirow{4}{*}{B} & \multirow{2}{*}{Val} & nnU-Net & 52.97$\pm$22.66 & 55.62$\pm$19.63 & 99.47$\pm$0.71 & 66.82$\pm$6.06 & 79.78$\pm$4.35\tabularnewline 
 &  & AGU-Net & 39.71$\pm$28.25 & 51.54$\pm$20.59 & 81.25$\pm$6.47 & 82.30$\pm$4.60 & 81.52$\pm$2.97\tabularnewline 
 & \multirow{2}{*}{Test} & nnU-Net & 59.19$\pm$20.49 & 61.61$\pm$16.72 & 98.04 & 80.65 & 88.50\tabularnewline 
 &  & AGU-Net & 43.76$\pm$27.61 & 53.14$\pm$20.23 & 84.31 & 87.76 & 86.00\tabularnewline 
\hline 
\multirow{4}{*}{C} & \multirow{2}{*}{Val} & nnU-Net & 52.43$\pm$22.45 & 54.72$\pm$19.77 & 99.81$\pm$0.35 & 63.70$\pm$6.68 & 77.58$\pm$5.01\tabularnewline 
 &  & AGU-Net & 37.43$\pm$28.69 & 51.09$\pm$20.49 & 79.29$\pm$10.08 & 84.70$\pm$3.23 & 81.32$\pm$4.43\tabularnewline 
 & \multirow{2}{*}{Test} & nnU-Net & 58.14$\pm$21.01 & 60.51$\pm$17.52 & 100.00 & 76.12 & 86.44\tabularnewline 
 &  & AGU-Net & 42.33$\pm$27.87 & 53.97$\pm$18.51 & 78.43 & 95.24 & 86.02\tabularnewline 
\hline 
\multirow{4}{*}{D} & \multirow{2}{*}{Val} & nnU-Net & 52.80$\pm$22.59 & 55.26$\pm$19.73 & 99.66$\pm$0.44 & 66.21$\pm$5.72 & 79.42$\pm$4.15\tabularnewline 
 &  & AGU-Net & 41.02$\pm$28.08 & 52.45$\pm$20.14 & 82.80$\pm$5.27 & 85.16$\pm$5.24 & 83.73$\pm$3.17\tabularnewline 
 & \multirow{2}{*}{Test} & nnU-Net & 58.05$\pm$22.74 & 60.42$\pm$19.61 & 100.00 & 79.69 & 88.70\tabularnewline 
 &  & AGU-Net & 40.84$\pm$28.62 & 52.07$\pm$20.96 & 78.43 & 93.02 & 85.11\tabularnewline 
\hline 
\multirow{4}{*}{E} & \multirow{2}{*}{Val} & nnU-Net & 53.61$\pm$22.57 & 55.81$\pm$19.97 & 100.00 & 63.86$\pm$6.93 & 77.73$\pm$5.19\tabularnewline 
 &  & AGU-Net & 39.44$\pm$27.05 & 48.89$\pm$20.92 & 85.61$\pm$4.83 & 84.58$\pm$3.39 & 84.91$\pm$1.59\tabularnewline 
 & \multirow{2}{*}{Test} & nnU-Net & 56.30$\pm$21.07 & 58.60$\pm$17.84 & 100.00 & 76.12 & 86.44\tabularnewline 
 &  & AGU-Net & 41.23$\pm$25.72 & 47.78$\pm$20.93 & 86.27 & 89.80 & 88.00\tabularnewline 
  \end{tabular}}}
  \label{tab:segmentation_results}
\end{table}

Segmentation performances across both architectures, for all input sequences combinations, and only for patients with residual tumor are summarized in Table~\ref{tab:segmentation_results}. 
For both architectures, the lowest average Dice score over the external test set was obtained with configuration A, indicating that solely using T1w-CE MR scans is insufficient for identifying post-operative residual tumor. The addition of the T1w scan as input (i.e., configuration B) provides at least a 5\% improvement in Dice scores over the test set for both architectures. This illustrates the additional value of the T1w sequence, presumably due to better distinction between blood and tumor.
The inclusion of the FLAIR scan in input configuration C slightly degraded the Dice score compared to input configuration B.
Finally, the inclusion of pre-operative data does not seem to improve the performance for any architecture, as the Dice scores for input configuration D are again slightly lower than for configuration B. Further addition of the FLAIR scan in input configuration E leads to a minor decrease in Dice scores compared to configuration D. For both architectures, input configuration B yielded the highest Dice scores on the test set. The highest Dice and true positive Dice scores were obtained with the nnU-Net architecture trained on input configuration B, with respectively 59\% and 61\% Dice on the validation and test sets. 
Overall, performances obtained across the test set are stable, in support of generalizability. Likewise, performances over the validation sets from the cross-validation protocol are consistent for inputs configurations B to E. The same results trends can be observed across both architectures for the true positive Dice, although slightly higher for the positive Dice using the nnU-Net architecture.

Looking at patient-wise performances, models trained with the nnU-Net architecture achieve perfect recall across all configurations for both the validation and test sets. Whereas the patch-wise strategy followed allows for segmenting smaller structures, the loose criterion to consider a network prediction as true positive further strengthens this aspect. Indeed, only a few correctly overlapping voxels between the prediction and the ground truth are needed for residual tumor to be considered satisfactorily identified patient-wise. Due to the full volume approach, models trained with AGU-Net generally struggle to identify small elements, as indicated by an overall around 80\% across the board.
Conversely, the opposite trend can be noticed in regards to patient-wise precision performance. Models trained with nnU-Net tend to perform more erroneous predictions as indicated by average precision scores below 70\%, whereas AGU-Net models tend to be more precise with precision scores up to 95\%.
Ultimately, F1-scores are very similar between models from both architectures across the different input configurations, as they combine recall and precision performances. 

\begin{figure}[!ht]
    \centering
    \includegraphics[width=\textwidth]{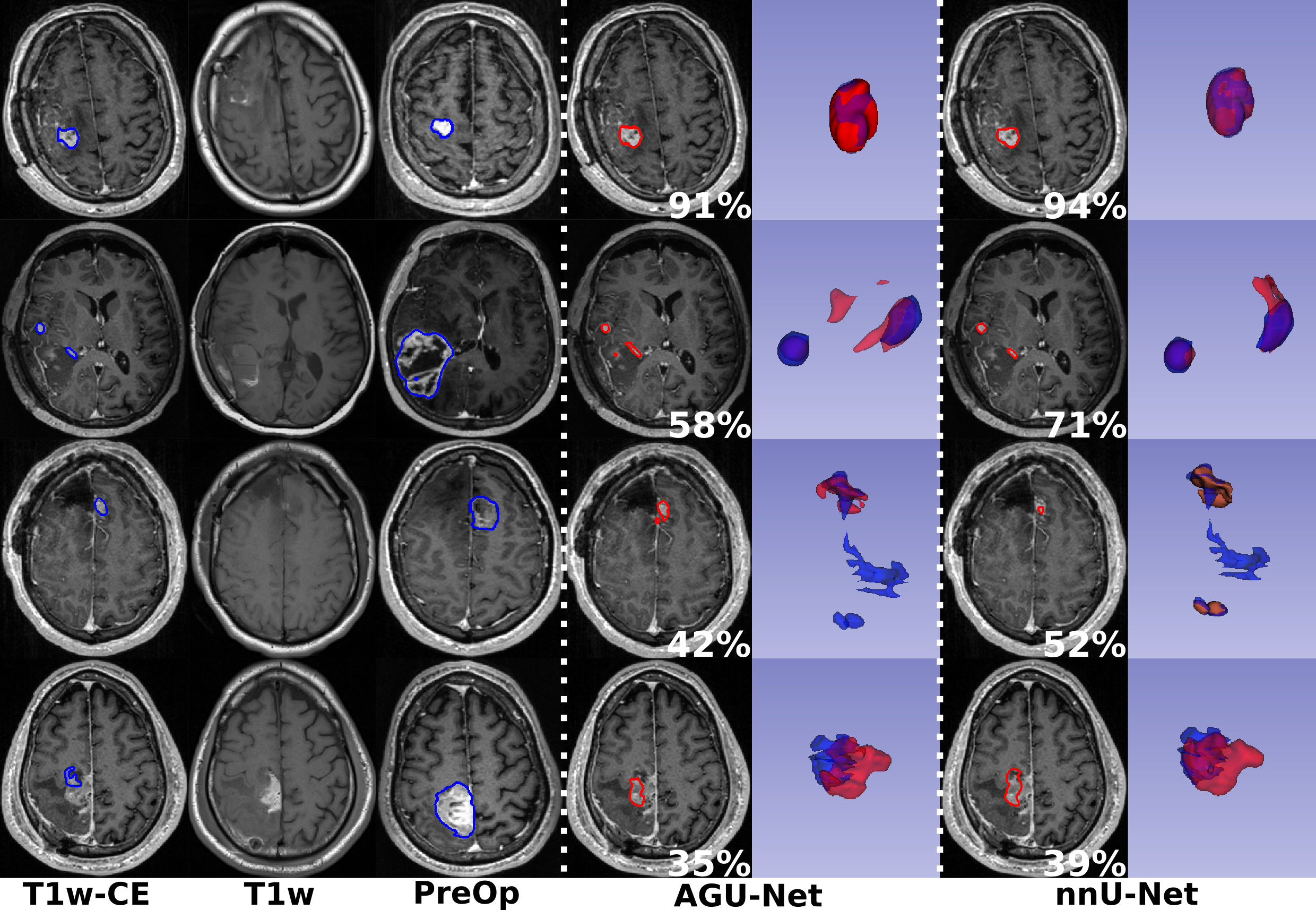}
    \caption{Segmentation comparison between the manual ground truth (in blue) and the binary predictions (in red) for the two architectures using configuration D, over the test set. One patient is featured per row, the patient-wise Dice is reported in white, and a 3D representation of the overlap is included (best viewed digitally and in color).}
    \label{fig:seg_results_illu}
\end{figure}

From the segmentation performances analysis, the best results have been obtained with the nnU-Net architecture using input configuration D.
Visual comparisons are provided in Figure~\ref{fig:seg_results_illu} between the two architectures using the best input configuration for some patients from the test set, one featured per row. In the top row, both models achieved excellent segmentation with a Dice score above 90\%. In the second row, a multifocal post-operative residual tumor case is featured whereby the AGU-Net model produced one false positive component as can be seen in red in the 3D representation. For the third row, a challenging multifocal and fragmented residual tumor case is displayed where both models failed to segment the largest component. Finally, in the last row, oversegmentation was  performed using both models leading to Dice scores below 40\%.

\subsection{Gross total resection classification performance study}

\begin{table}[!ht]
 \caption{Gross total resection versus residual tumor classification performances for both architectures, all input configurations, and over the validation and test sets.}
  \centering
  {\adjustbox{max width=\textwidth}{
  \begin{tabular}{lll|ccc}
    \multirow{2}{*}{\bfseries Exp.} & \multirow{2}{*}{\bfseries Data} & \multirow{2}{*}{\bfseries Arch.} & 
    \multicolumn{3}{c}{\bfseries Patient-wise} \tabularnewline
        & & & \bfseries Sensitivity & \bfseries Specificity & \bfseries bAcc\tabularnewline
\hline 
\multirow{4}{*}{A} & \multirow{2}{*}{Val} & nnU-Net & 99.81$\pm$0.35 & 2.53$\pm$2.21 & 51.17$\pm$1.22\tabularnewline 
 &  & AGU-Net & 79.70$\pm$6.69 & 68.01$\pm$10.43 & 73.86$\pm$4.94\tabularnewline 
 & \multirow{2}{*}{Test} & nnU-Net & 100.00 & 4.55 & 52.27\tabularnewline 
 &  & AGU-Net & 84.31 & 63.64 & 73.98\tabularnewline 
\hline 
\multirow{4}{*}{B} & \multirow{2}{*}{Val} & nnU-Net & 99.47$\pm$0.71 & 18.04$\pm$4.41 & 58.75$\pm$2.30\tabularnewline 
 &  & AGU-Net & 81.25$\pm$6.47 & 71.01$\pm$5.36 & 76.13$\pm$4.12\tabularnewline 
 & \multirow{2}{*}{Test} & nnU-Net & 98.04 & 45.45 & 71.75\tabularnewline 
 &  & AGU-Net & 84.31 & 72.73 & 78.52\tabularnewline 
\hline 
\multirow{4}{*}{C} & \multirow{2}{*}{Val} & nnU-Net & 99.81$\pm$0.35 & 5.64$\pm$3.44 & 52.73$\pm$1.76\tabularnewline 
 &  & AGU-Net & 79.29$\pm$10.08 & 74.00$\pm$11.13 & 76.64$\pm$4.87\tabularnewline 
 & \multirow{2}{*}{Test} & nnU-Net & 100.00 & 27.27 & 63.64\tabularnewline 
 &  & AGU-Net & 78.43 & 90.91 & 84.67\tabularnewline 
\hline 
\multirow{4}{*}{D} & \multirow{2}{*}{Val} & nnU-Net & 99.66$\pm$0.44 & 15.28$\pm$6.85 & 57.47$\pm$3.55\tabularnewline 
 &  & AGU-Net & 82.80$\pm$5.27 & 73.00$\pm$14.63 & 77.90$\pm$6.44\tabularnewline 
 & \multirow{2}{*}{Test} & nnU-Net & 100.00 & 40.91 & 70.45\tabularnewline 
 &  & AGU-Net & 78.43 & 86.36 & 82.40\tabularnewline 
\hline 
\multirow{4}{*}{E} & \multirow{2}{*}{Val} & nnU-Net & 100.00 & 6.12$\pm$4.30 & 53.06$\pm$2.15\tabularnewline 
 &  & AGU-Net & 85.61$\pm$4.83 & 72.63$\pm$9.39 & 79.12$\pm$4.60\tabularnewline 
 & \multirow{2}{*}{Test} & nnU-Net & 100.00 & 27.27 & 63.64\tabularnewline 
 &  & AGU-Net & 86.27 & 77.27 & 81.77\tabularnewline 
  \end{tabular}}}
  \label{tab:classification_results}
\end{table}

Classification performances between patients with residual tumor and gross total resections, across both architectures and for all input configurations, are reported in Table~\ref{tab:classification_results}. The first noticeable result is the overall tendency of the nnU-Net architecture to oversegment, resulting in a perfect recall over both the test set and validation set, for a really poor specificity often below 30\%. Overall, nnU-Net achieves balanced accuracy scores barely above 0.5 for all input configurations, which means the classification performance is only slightly better compared to the average score of random guessing (i.e., 0.5). Conversely, models trained with the AGU-Net architecture are more conservative leading to higher specificity scores, up to 90\% for input configuration C, and reasonably high recall/sensitivity values above 80\%.
In contrast to segmentation performances, the successive addition of MR scans within the input configuration lead to improved classification performances for both architectures. One apparent difference is the added value of the FLAIR sequence with the AGU-Net architecture, further increasing the specificity and balanced accuracy, unlike performances with the nnU-Net architecture.

From the classification performances analysis, the best results on the test set according to the balanced accuracy have been obtained with the AGU-Net architecture using input configuration C. However, the best results on the validation sets are achieved with input configuration E. In a clinical scenario, a high sensitivity has higher priority than a high specificity, as long as the trade-off is reasonable. AGU-Net trained with input configuration E is therefore the preferred model for classification. This configuration achieves the highest sensitivity for all input configurations while still achieving a reasonable specificity, higher than configurations A and B. 

\subsection{Inter-rater variability study} 

\begin{figure}[!ht]
\centering
\includegraphics[width=\textwidth]{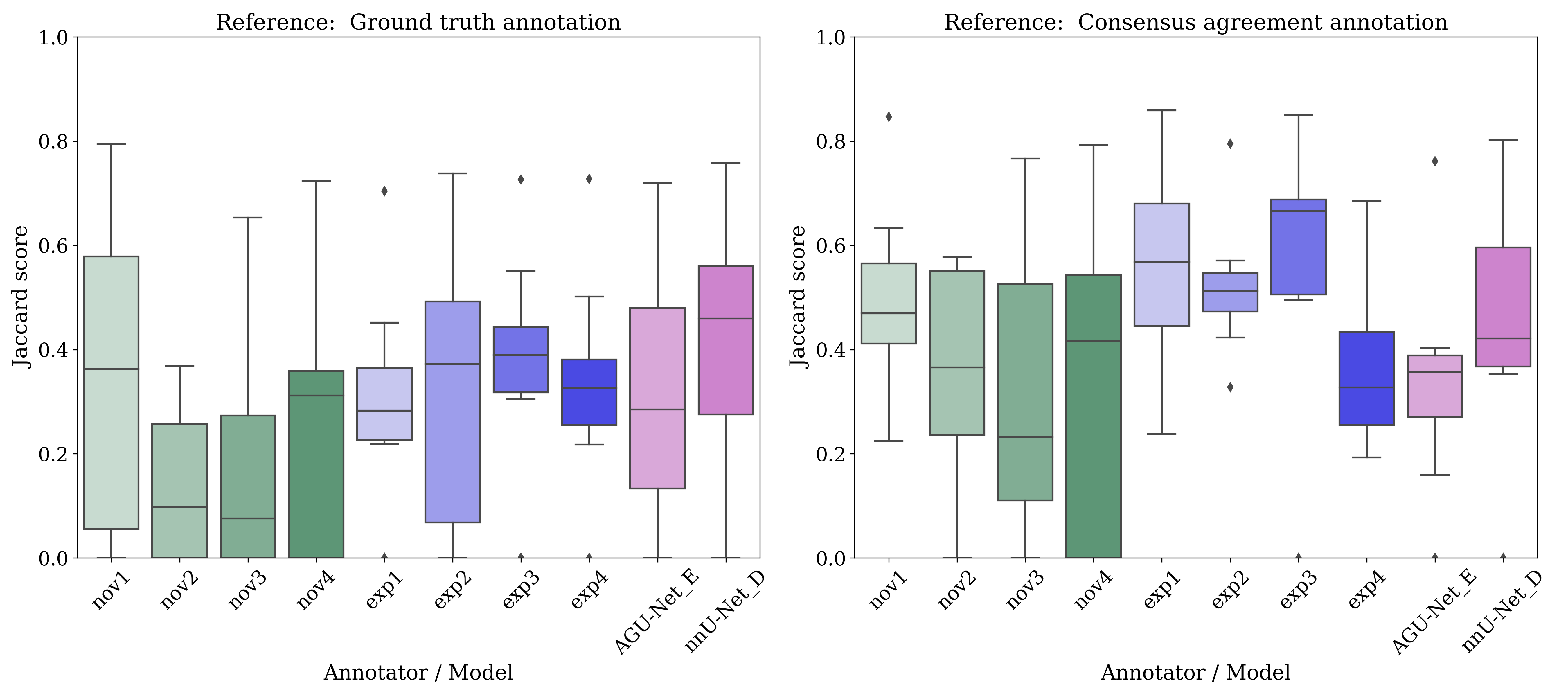}
\caption{Inter-rater Jaccard score variability over a subset of the AMS cohort. To the left, the ground truth annotation used for training served as segmentation of reference. To the right, the reference segmentation was a consensus agreement between annotations from all raters.}
\label{fig:interrater-illu}
\end{figure}

For the 20 patients constituting the inter-rater variability dataset, a comparison of the Jaccard scores, obtained between each rater and the best model from each architecture, are reported in Figure~\ref{fig:interrater-illu}. As all configurations B-E yielded very similar segmentation performance scores, the selected best configuration for each architecture were the models that produced the best trade-off between sensitivity and specificity for the classification task. For the nnU-Net, configuration D was selected as the best configuration as this model achieved the highest specificity out of all the trained nnU-Net models, and for the AGU-Net, configuration E was selected as this model yielded the highest sensitivity while maintaining a reasonable specificity. Using the ground truth annotation from the test dataset as a reference segmentation, both architectures achieved Jaccard scores within a variability range very similar to that of the novice and expert annotators. With the consensus agreement annotation as a reference segmentation, the AGU-Net model achieved slightly poorer Jaccard scores than the majority of the expert human raters but remained within the variability range of the novice annotators. The nnU-Net model achieved scores similar to the variability range of the expert raters, also when compared to the consensus agreement annotation.

\section{Discussion}
In this multicenter study, the feasibility of post-operative residual tumor segmentation with deep neural networks was assessed. Two state-of-the-art architectures for pre-operative glioblastoma segmentation were compared: nnU-Net and AGU-Net. Both architectures were trained on five different combinations of early post-operative and pre-operative MR scans as input, and benchmarked in terms of segmentation and classification performances compared with manual rating. The main finding is that automatic segmentation performances are comparable to human rater performance on real world MRI scans, requiring early post-operative T1w-CE and T1w MRI scans only. In addition, the trained automated models have shown promising ability to classify patients who underwent gross total resection from patients exhibiting post-operative residual tumor.

The multimodal and multicentric dataset in this study is the largest cohort used for the task of early post-operative glioblastoma segmentation, with a total of 956 patients. Regarding the dataset curation, our strict inclusion criteria required availability of all four MR scans as input (i.e., post-operative T1w-CE, T1w, FLAIR, and pre-operative T1w-CE) for each patient. Whereas this decision was motivated by a simpler method design, approximately 150 patients were excluded as one or more MR scans were missing. 
A relaxation of the inclusion criteria would increase the size of the dataset, and open the possibility to generate a more diverse set of input MR scans, including for example T2-weighted images. Ideally, the trained methods should be able to deal with a sparse set of inputs, where one or more MR scans are missing. The trained models should be used off the shelf, by replacing missing sequences with empty volumes, synthetically generated sequences, or allowing missing inputs using sparsified learning techniques~\cite{grovik2021handling}.

In their ability to segment post-operative tumor, nnU-Net and AGU-Net exhibit strengths and weaknesses inherent to their design. Through a patch-wise approach, nnU-Net models are able to segment relatively small structures, having access to more fine-grained details from the use of MR scans close to their initial resolution. Considering the relatively small volumes and fragmented state for residual tumors, nnU-Net models are able to achieve higher Dice score and recall performances. On the other hand, models trained using the AGU-Net approach are following a full volume approach, largely downsampling the input MR scans, hindering the ability for detecting smaller structures. However, such models appear to be more conservative in their predictions, hence heavily reducing the amount of false positives enabling to reach high precision performances.
Regarding the different input configurations, the biggest impact on segmentation performances comes from combining EPMR T1w-CE and T1w scans, which corresponds to the favored approach as well in clinical practice. The inclusion of additional MR sequences seems to add little to segmentation performances. Adapting the convolution blocks, filter sizes, or other elements of the architectures might be needed for letting the number of trainable parameters to evolve according to the number of inputs, instead of a fixed amount of parameters.

The validation studies described in this article served the two purposes of investigating the predictive ability and capacity to generalize of the trained models. This is obtained through the use of a unique test set, and equally distributed hospital-stratified validation sets. Our selection for a specific hold-out hospital as a test set was based on the availability of manual annotations from multiple raters, allowing to perform, in addition, an inter-rater variability study.
Regarding the computation of the reported metrics, the rationale for only including the true positive patients in the segmentation performances lies in the Dice score computation itself. Indeed, cases with a GTR preclude calculation of a Dice score. Therefore, the validation studies include a separate experiment to classify patients into those with a GTR and those with residual tumor.

The inter-rater variability study demonstrated that residual tumor segmentation performance is on par with the average human expert annotator performance, when evaluated against an independent ground truth segmentation. Even when evaluated against the consensus agreement annotation, which is by definition biased towards each of the human annotators included in the study, the best segmentation model achieves scores similar to the individual expert annotations, and still outperforms the novice annotators. The consensus agreement annotation based on a majority voting scheme over all annotations from the eight different annotators should be considered the gold standard for defining the residual tumor. However, this is not achievable in a real-world clinical scenario, where even an exact delineation of the tumor remnant from one human annotator is rarely performed. The proposed automatic method for residual tumor segmentation should thus be considered an acceptable alternative to the current standard practice for evaluating the tumor remnant after surgery, as the average performance of the method lies within the variability range of individual expert annotators.
Such segmentation performances are even achieved with the exclusive use of post-operative MR sequences as model inputs (T1w-CE, T1w, and FLAIR), whereas the addition of pre-operative information (pre-operative T1w-CE and label) retains the model performance on similar levels. Thus, in clinical practice, our trained models could be deployed even in the absence of pre-operative scans, as long as at least the T1w-CE and T1w post-operative sequences are available, to establish an automated and relatively fast method for the segmentation task. On a second level, the output segmentation masks can be used to differentiate between patients with remnant tumor after surgery and gross total resection patients, with increasing balanced accuracy performance as more sequences are added to the model inputs. 
Our early post-operative glioblastoma segmentation models have been made freely available in the Raidionics environment\footnote{\url{https://github.com/raidionics}}.

In spite of promising reported performances, the task of early post-operative glioblastoma segmentation is far from accomplished. The full extent of residual tumor, often very fragmented around the resection cavity, is never wholly captured. In future work, the pre-operative MR scans and tumor location should be better leveraged as the residual tumor is bound to lie in its vicinity. Focusing the search solely within a region of interest might help retaining a higher image resolution, for better segmentation of small structures.
Nevertheless, competitive pre- and post-operative glioblastoma segmentation models are now publicly available, opening the door to clinically-oriented validation studies. Assuming a positive outcome, the use of automatic models and methods would be highly beneficial in a clinical setting to collect parameters currently obtained through eyeballing or diameter estimation, hence yielding reproducible and deterministic significance.

\section{Conclusion}
In this study, two state-of-the-art neural network architectures for glioblastoma segmentation were trained and thoroughly validated on a large cohort of 956 patients. Automatic segmentation performances are on par with human rater performance on real world MRI scans, requiring early post-operative T1w-CE and T1w MRI scans only. In addition, the presented models have shown promising readiness for automatically distinguishing between patients who underwent gross total resection, and patients with residual tumor. The prognostic value of the automated method should be assessed in future studies.

\bibliography{article}

\section*{Acknowledgements}
Data were processed in digital labs at HUNT Cloud, Norwegian University of Science and Technology, Trondheim, Norway.
FB is supported by the National Institute for Health Research (NIHR) biomedical research centre at UCLH. The PICTURE project is sponsored by an unrestricted grant of Stichting Hanarth fonds, “Machine learning for better neurosurgical decisions in patients with glioblastoma”; a grant for public-private partnerships (Amsterdam UMC PPP-grant) sponsored by the Dutch government (Ministry of Economic Affairs) through the Rijksdienst voor Ondernemend Nederland (RVO) and Topsector Life Sciences and Health (LSH), “Picturing predictions for patients with brain tumors”; a grant from the Innovative Medical Devices Initiative program, project number 10-10400-96-14003; The Netherlands Organisation for Scientific Research (NWO), 2020.027; a grant from the Dutch Cancer Society, VU2014-7113 and the Anita Veldman foundation, CCA2018-2-17. R.H.H. is supported by a grant from The Research Council of Norway, grant number 323339. D.B., I.R., and O.S. are partly funded by the Norwegian National Research Center for Minimally Invasive and Image-Guided Diagnostics and Therapy.

\section*{Author contributions statement}
Conceptualization: R.H.H., A.F., D.B., R.E., I.R., O.S., and P.C.D.W.H.; Data curation: I.K., H.A., F.B., L.B., M.B., T.D., M.N., J.F., S.H.J., A.I., B.K., R.T., E.M., D.M., P.R., M.R., L.S., T.S., T.A., M.W., G.W., M.W., A.Z., P.M., A.J., O.S., P.C.D.W.H.; Methodology, Investigation, Formal analysis, and Validation: R.H.H., A.F., D.B., and R.E.; Funding acquisition and project administration: I.R., O.S, and P.C.D.W.H.; Software: D.B., A.P., R.H.H.; Writing - original draft: R.H.H., A.F., D.B., A.P., R.E., I.R., O.S., and P.C.D.W.H.; Writing - review \& editing: all authors.

\section*{Additional information}
\textbf{Accession codes}
The best trained AGU-Net model can be accessed at~\url{https://github.com/raidionics/Raidionics-models/releases/tag/1.2.0}. The best trained nnU-Net model can be accessed at~\url{https://gitlab.com/picture-production/picture-nnunet-package/tree/0.3.7}.
The source code used for computing the metrics can be accessed at~\url{https://github.com/dbouget/validation_metrics_computation}.
Inference on new patients can be performed using the Raidionics software which is openly available at~\url{https://github.com/raidionics/Raidionics}.

\textbf{Competing interests} The authors declare no conflict of interest. The funders had no role in the design of the study; in the collection, analyses, or interpretation of data; in the writing of the manuscript; nor in the decision to publish the results.


\end{document}